\documentclass[twocolumn,prl,preprintnumbers,amsmath,amssymb,superscriptaddress,floatfix]{revtex4}

\usepackage{latexsym}
\usepackage{epsfig}
\usepackage{graphicx}
\usepackage{textcomp}
\usepackage{color}
\usepackage{mathtools}

\begin{document}



\title{Robust Measurement for the Discrimination of Binary Coherent States}

\author{M. T. DiMario}

\affiliation{Center for Quantum Information and Control, Department of Physics and Astronomy, University of New Mexico,
Albuquerque, New Mexico 87131}

\author{F. E. Becerra}

\affiliation{Center for Quantum Information and Control, Department of Physics and Astronomy, University of New Mexico,
Albuquerque, New Mexico 87131}
\email{fbecerra@umd.edu}

%

\begin{abstract}

The discrimination of two nonorthogonal states is a fundamental element for secure and efficient communication. Quantum measurements of nonorthogonal coherent states can enhance information transfer beyond the limits of conventional technologies. We demonstrate a strategy for binary state discrimination based on optimized single-shot measurements with photon number resolving (PNR) detection with finite number resolution. This strategy enables a high degree of robustness to noise and imperfections while being scalable to high rates and in principle allows for surpassing the quantum noise limit (QNL) in practical situations. These features make the strategy inherently compatible with high-bandwidth communication and quantum information applications, providing advantages over the QNL under realistic conditions.
\end{abstract}

\maketitle

The realization of quantum technologies that can provide advantages over conventional ones is central in quantum information. Discrimination measurements of nonorthogonal coherent states is an integral part of quantum communication protocols \cite{bennett84, bennett92, huttner95,grosshans02, silberhorn02,grosshans03, gisin02, sych10, bergou04, giovannetti04}, and can assist quantum-state preparation and detection \cite{munro05, vanloock11}, entanglement generation \cite{vanloock08}, and computing \cite{nemoto04, ralph03}. Moreover, optimized measurements for the discrimination of coherent states with different phases can achieve sensitivities beyond the ideal limit for conventional technologies, called the quantum noise limit (QNL), and allow for approaching the ultimate limits of information transfer \cite{guha11, rosati16, klimek16}.
However, unavoidable noise and imperfections in realistic situations compromise the sensitivity performance of these optimized strategies, and put in question their advantages over conventional technologies in real-world applications. Furthermore, in order to be realistic alternatives to conventional technologies, these new measurement strategies should allow for scalability and be compatible with high-bandwidth communications, while requiring low complexity for their implementation.

Discrimination strategies based on complex feedback operations that approach the ultimate bound for coherent state discrimination, the Helstrom bound \cite{helstrom76}, have been proposed \cite{dolinar73, bondurant93, becerra11, sych10, nair14, muller12, muller15, izumi13, burenkov18} and experimentally demonstrated \cite{cook07, becerra13, becerra15,ferdinand17}. However, these strategies are not readily compatible with current communication technologies, since feedback operations limit the overall achievable communication bandwidth. On the other hand, discrimination strategies based on single-shot measurements that do not require feedback \cite{kennedy72, taeoka08, wittman08, tsujino11, izumi12, dimario18} can still provide advantages over the QNL, while being compatible with high-bandwidth communications. However, these strategies are not inherently robust against noise and imperfections of realistic communication channels, which has limited their performance below the QNL to very small power ranges \cite{wittman08, tsujino11}.

Here we investigate and experimentally demonstrate a strategy for the discrimination of two nonorthogonal states that combines the simplicity of optimized single-shot measurements with photon number resolution to provide robustness against noise. Photon number resolving (PNR) detection was proposed \cite{wittmann10b} and experimentally demonstrated \cite{wittmann10} to enable probabilistic discrimination of two nonorthogonal states based on measurements with postselection outperforming a post-selected homodyne measurement. Here we demonstrate that the use of PNR detection enables robustness against noise and imperfections in deterministic measurements for minimum-error discrimination. This strategy is compatible with high-bandwidth communication and information technologies, and provides advantages over the homodyne limit (the QNL) at arbitrary input powers, even in the presence of realistic noise and imperfections. Unlike previous single-shot strategies of two states with on-off detection for minimum-error discrimination \cite{kennedy72, wittman08, taeoka08}, our strategy uses PNR detection, which increases the number of possible outcomes from photon detections. This extension provides a dramatic increase in robustness to experimental imperfections analogous to feedback strategies for multiple states \cite{becerra15}, but retains the simplicity of single-shot measurements.
Our proof-of-principle experimental demonstration of the generalized strategy outperforms the QNL adjusted for our system detection efficiency. We observe that increasing the photon number resolution of the detector allows for extending discrimination below the QNL to higher input power levels in situations with noise and imperfections.

\begin{figure}[!tb]
\centering\includegraphics[width=8.8 cm]{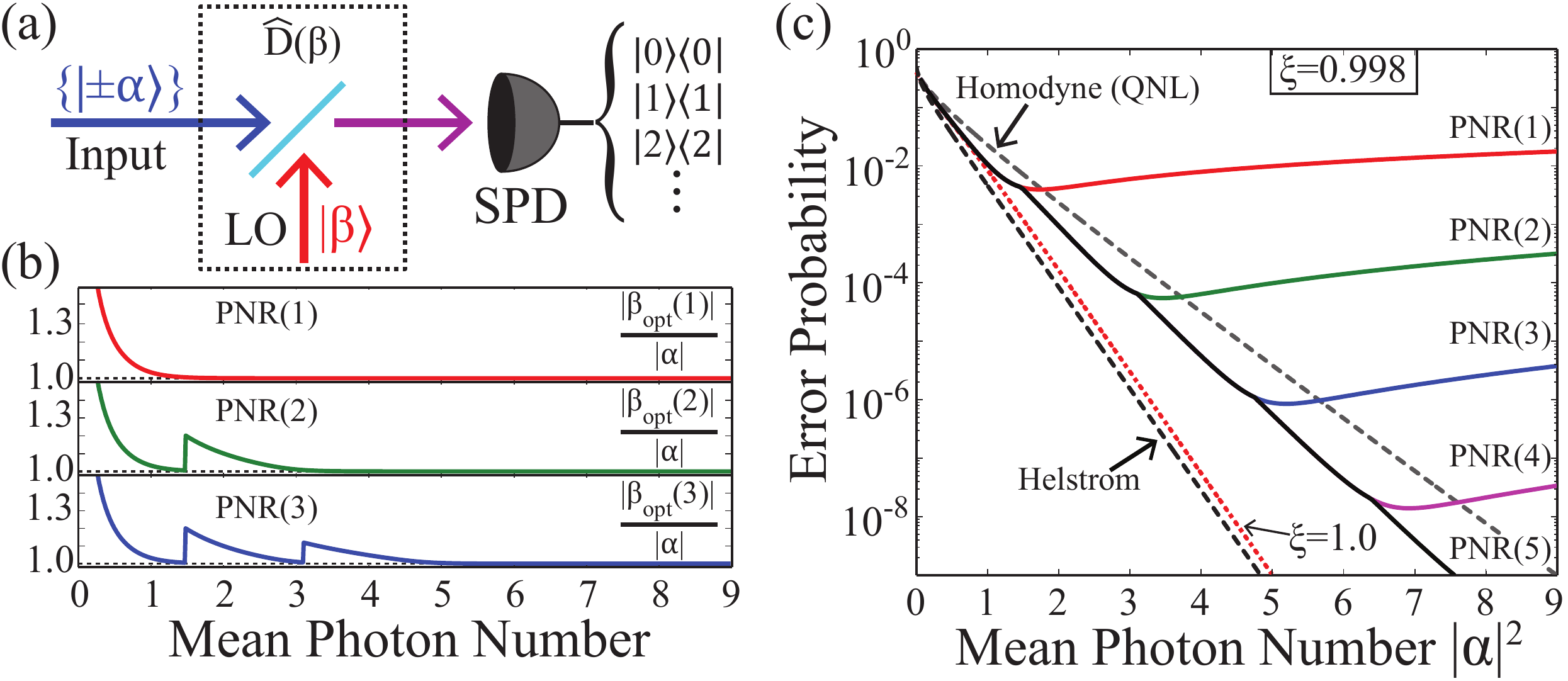}
\caption{\label{A} \textbf{Robust PNR optimized discrimination strategy.} (a) 
The input state $|\psi\rangle \in \{|\alpha \rangle, |-\alpha \rangle \}$ is displaced to $\hat{D}(\beta)|\psi\rangle$ using a strong local oscillator (LO) field. A photon number resolving (PNR) detector with finite photon number resolution PNR($m$) is used for the discrimination of the input state $|\psi\rangle$ in the presence of noise and imperfections, characterized by the reduction of visibility ($\xi$) of the displacement operation. (b) Optimal displacement ratios $|\beta_\mathrm{opt}(m)|/|\alpha|$ that minimize the probability of error for strategies with PNR(1), PNR(2), and PNR(3) with $\xi=0.998$. (c) Probability of error for the discrimination of two coherent states $\{|\alpha \rangle, |-\alpha \rangle \}$ with strategies with different number resolutions PNR($m$) with detection efficiency $\eta = 1.0$ and $\xi = 0.998$. Higher photon number resolution extends discrimination below the QNL (grey, dashed) at high powers in the presence of noise and imperfections. The Helstrom bound (black, dashed) and the expected error for the noiseless PNR(1) strategy (red, dotted) with visibility $\xi = 1.0$ are shown for reference.}
\end{figure}

\emph{Robust measurement.\textemdash}
Figure 1(a) shows the robust PNR optimized discrimination strategy. An input state $|\psi \rangle \in \{|-\alpha \rangle , |\alpha\rangle\}$, where $\alpha$ is real and positive, is displaced in phase space to $\hat{D}(\beta)|\psi \rangle$, where the displacement $\hat{D}(\beta)$ is implemented with a strong local oscillator (LO) field and a high-transmittance beam splitter \cite{paris96}. The displaced state is then detected by a PNR detector with outcomes corresponding to projections onto Fock states: $\hat{\Pi}_{n}=|n\rangle\langle n|$. The number of photons that a realistic PNR detector can resolve before acting as a threshold detector is referred to as the photon number resolution PNR($m$), where $m$ represents the maximum number of resolved photons \cite{becerra15}. For example, a PNR detector with PNR(3) has four measurement outcomes corresponding to $\{0, 1, 2, 3+\}$ photons. Here 3+ refers to the number of photons detected being three or greater.

The discrimination strategy uses the maximum $a$ $posteriori$ probability (MAP) criterion to estimate the input state based on the detection outcomes and the PNR resolution \emph{m}. Given the number of detected photons ($n$), the number resolution ($m$) of the detector, and the displacement field $\beta$, the strategy's decision about the input state corresponds to the state with the highest conditional posterior probability $P(\pm\alpha | \beta,n,m)$, which is obtained through Bayes' rule:

\begin{equation}
P(\pm\alpha |\beta, n, m)P(n,m) = P(n |\pm\alpha,\beta, m)P(\pm\alpha).
\end{equation}

Here $P(\pm\alpha)$ is the prior probability of input state $|\pm\alpha\rangle$, which is set to 0.5 for our experiment. $P(n | \pm\alpha, \beta, m)$ is the conditional probability of detecting $n$ photons given the input state is displaced by $\beta$ with a strategy with PNR($m$), and $P(n,m)$ is the probability of detecting $n$ photons: $P(n,m) =  (P(n|-\alpha,\beta,m)+P(n|\alpha,\beta,m))/2$. Using this strategy, the probability of error for a discrimination measurement with PNR(\emph{m}) becomes (see Supplemental Material \cite{Supplemental}):

\begin{equation}
P_\mathrm{E}(\alpha, \beta, m) = 1 - \frac{1}{2}\sum_{k=0}^{m} max(\{P(k|\pm\alpha, \beta, m)\})
\end{equation}

Here the conditional probabilities $P(k | \pm\alpha, \beta, m)$ of detecting $k$ photons given the input state is displaced by $\beta$ are given by the Poisson probabilities:

\begin{align}
&P(k|\pm\alpha,\beta, m) =
\begin{dcases}
      \frac{\langle n \rangle_{\pm}^{k}}{k!}e^{-\langle n \rangle_{\pm}}  , & k<m \\
      1-\sum_{l=0}^{m-1} \frac{\langle n \rangle_{\pm}^{l}}{l!}e^{-\langle n \rangle_{\pm}} , & k=m \\
 \end{dcases}
 \\
 \nonumber
 \\
 &\langle n \rangle_{\pm} = |\langle \beta | \pm\alpha\rangle|^{2} = |\alpha|^{2}+|\beta|^{2} \pm 2\xi|\alpha||\beta|
\end{align}
\\
where $\xi$ is the interference visibility of the displacement operation. The reduction of visibility quantifies the noise and imperfections that affect the discrimination process \cite{becerra15}, and has limited the performance of binary discrimination measurements below the QNL to very small power ranges \cite{cook07, wittman08, tsujino11}.

For a given discrimination strategy with PNR($m$), the amplitude $|\beta|$ of the displacement operation can then be optimized to minimize the probability of error $P_\mathrm{E}(\alpha, \beta, m)$ for a fixed mean-photon number of the input state $|\alpha|^{2}$:
\begin{equation}
\label{optimPe}
{\frac{\partial P_\mathrm{E}(\alpha, \beta, m)}{\partial \beta}}\Big|_{\beta_\mathrm{opt}(m)}=0.
\end{equation}
This optimization results in discrimination strategies which allow for surpassing the QNL at high input powers.

Figure 1(b) shows the ratio of the optimal displacement amplitude $|\beta_\mathrm{opt}(m)|$ to the amplitude of the input state $|\alpha|$, $|\beta_\mathrm{opt}(m)|/|\alpha|$ for strategies with PNR(1), PNR(2), and PNR(3) with detection efficiency $\eta=1$ and a level of noise and imperfections characterized by a reduced visibility of $\xi=0.998$. In general the optimal displacement ratios converge to 1 as $|\alpha|^{2}$ increases, which corresponds to displacing the input state to the vacuum state. However, as the photon number resolution $m$ is increased, the optimal displacements show sharp jumps at $|\alpha|^{2} \approx$ 1.5 and 3. These jumps come from solving the minimization in Eq. (5) 
 for strategies with different $m$ at these points, with $m-1$ jumps for a strategy with PNR($m$). See Supplemental Material \cite{Supplemental}.

Figure 1(c) shows the error probabilities (solid lines) for strategies with PNR(1)-PNR(5) with $\eta=1$ and $\xi=0.998$, together with the homodyne limit at the QNL (dashed grey line) given by:
\begin{equation}
P_\mathrm{hom}=\frac{1}{2}(1-\mathrm{erf}(\sqrt{2}\alpha)),
\label{optimPe}
\end{equation}
the Helstrom bound (dashed black line) given by:
\begin{equation}
P_\mathrm{hels}=\frac{1}{2}[1-\sqrt{1-\exp(-4|\alpha|^2)}],
\label{optimPe}
\end{equation}
and an ideal PNR(1) strategy with $\xi=1$ (dotted red line).
We observe that the points where jumps in optimal displacement ratio take place correspond to points at which a certain PNR($m$) strategy starts to degrade and splits off from the higher PNR($m$) strategies. This degradation in performance is caused by noise and imperfections characterized by the non-ideal visibility of the displacement operation.
We note that while the ideal PNR(1) measurement performs very close to the Helstrom bound, non-ideal visibility has a dramatic effect on its performance, which severely limits the power range at which a strategy with PNR(1) can surpass the QNL \cite{wittman08, tsujino11}. On the other hand, strategies with PNR detection provide robustness to noise and imperfections resulting in non-ideal visibility, and can extend discrimination below the QNL to higher powers as $m$ is increased. This shows that single-shot PNR receivers can beat the QNL under realistic conditions, mitigating the effects of noise and imperfections without requiring feedback operations, as long as the photon number resolution of the detector is high enough.

\begin{figure}[!tbp]
\centering\includegraphics[width=8.5 cm]{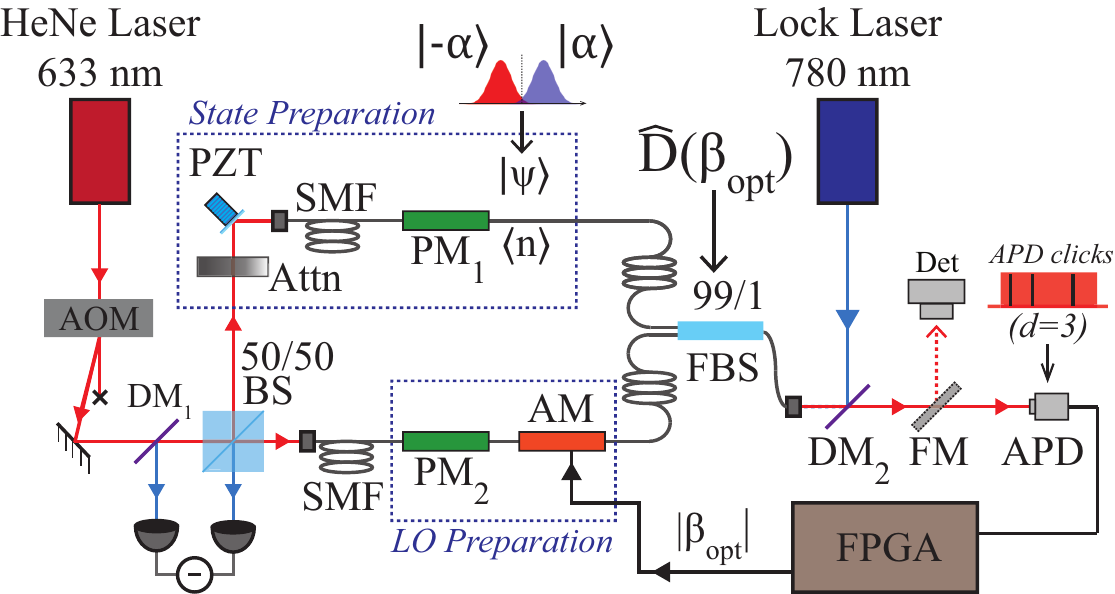}
\caption{\label{A} \textbf{Experimental Setup.} Coherent-state pulses from a HeNe laser and an acousto-optic modulator (AOM) are incident in an unbalanced Mach-Zehnder interferometer. Fiber-coupled phase modulator PM$_{1}$ prepares the input state $|\psi \rangle \in \{|-\alpha\rangle, |\alpha \rangle\}$. Phase (PM$_{2}$) and amplitude (AM) modulators prepare the optimized displacement field $\beta_\mathrm{opt}(m)$. A 99/1 fiber beam splitter (FBS) implements the optimal displacement operation $\hat{D}(\beta_\mathrm{opt}(m))$ for a given strategy with specific PNR($m$). An avalanche photodiode (APD) is used as a PNR detector and a field-programmable gate array (FPGA) collects the detector outcomes and controls the optimized displacements (see text for details). (DM), dichroic mirror; (Att), attenuator; (SMF), single-mode fiber; (DM), flip mirror; (DD), differential detector; (PZT), piezo; (Det), calibrated detector; (BS) beam splitter.
}\label{ExpFig}
\end{figure}

\emph{Experimental demonstration.\textemdash}
Figure \ref{ExpFig} shows the experimental setup for the demonstration of the optimized discrimination strategy with PNR detection. A HeNe laser at 633~nm and an acousto-optic modulator (AOM) prepare 26 $\mu$s light pulses at a rate of 11.7 kHz, which are sent to an interferometric setup. A fiber-coupled phase modulator PM$_{1}$ and an attenuator prepare the input state $|\psi\rangle$ with a phase of $\phi=0$ or $\phi=\pi$, and with a mean photon number $|\alpha|^2$ calibrated with a transfer standard detector (Det) \cite{gentile96}. A phase modulator PM$_{2}$ and an amplitude modulator AM prepare the optimal displacement field $\beta_\mathrm{opt}(m)$ for a specific strategy with specific resolution PNR($m$) from Eq. (\ref{optimPe}). The AM is controlled by an 8-bit register from a field-programmable gate array (FPGA) which is converted into an analog voltage with a digital-to-analog converter (DAC), allowing preparation of the amplitude of $\beta_\mathrm{opt}(m)$ with less than 1$\%$ error \cite{ferdinand17}. We estimate the error in the relative phase of the LO and the input state to be about 0.05 radians by observing the interference between these fields used for phase calibration \cite{becerra11}. The optimized displacement operation $\hat{D}(\beta_\mathrm{opt}(m))$ is performed with a 99/1 fiber beam splitter (FBS). The displaced state $\hat{D}(\beta_\mathrm{opt})|\pm \alpha \rangle$ is then detected by an avalanche photodiode (APD), and the detection outcomes are collected by the FPGA. We use the APD as a number resolving detector \cite{becerra15, wittmann10, banaszek99b} with a nonzero probability of afterpulsing $P_\mathrm{AP}$ (see Supplemental Material for details).

\begin{figure}[!bp]
\centering\includegraphics[width=8.5 cm]{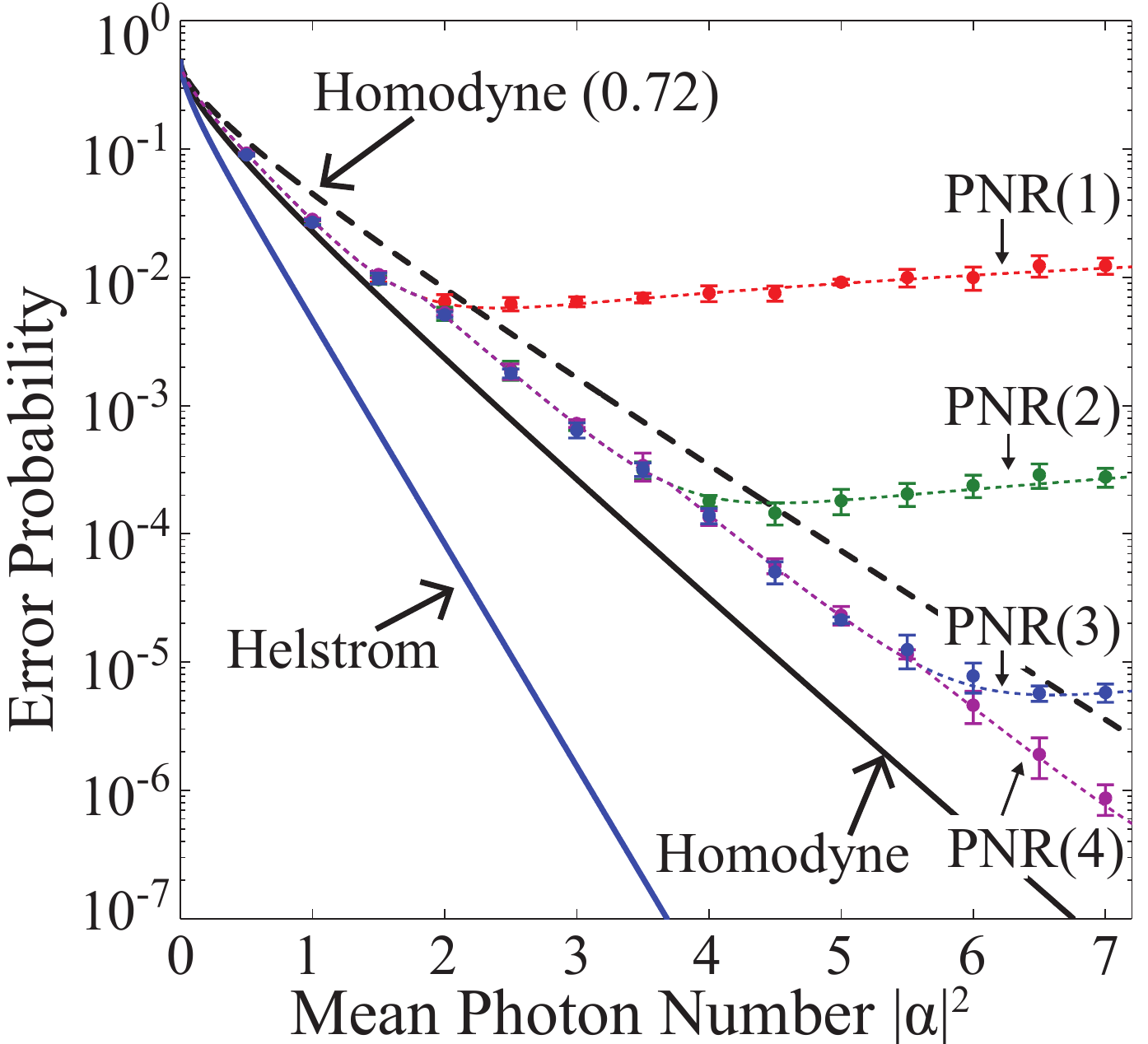}
\caption{\textbf{Experimental Results.} Experimental error probability for strategies with PNR($m$) for $m=1,2,3,4$ (solid circles) with error bars representing 1-$\sigma$ statistical standard deviation. Each data point is the result of five experimental runs for each mean photon number and PNR($m$) (see Supplemental Material for details). Also shown are the Helstrom bound (solid blue line), the QNL given by a perfect homodyne measurement (solid black line) and a homodyne measurement adjusted for our system detection efficiency of $\eta = 0.72(1)$ (dashed black line). The theoretical predictions with $\eta = 0.72$, $\xi = 0.998$, dark count rate $\nu = 3.6\times10^{-3}$, and afterpulsing probability $P_\mathrm{AP} =1.10\times10^{-2}$ (colored dashed lines) show very good agreement with the experimental observations.}
\label{ExpResults}
\end{figure}

\begin{figure*}[!t]
\centering
\includegraphics[width=1\textwidth]{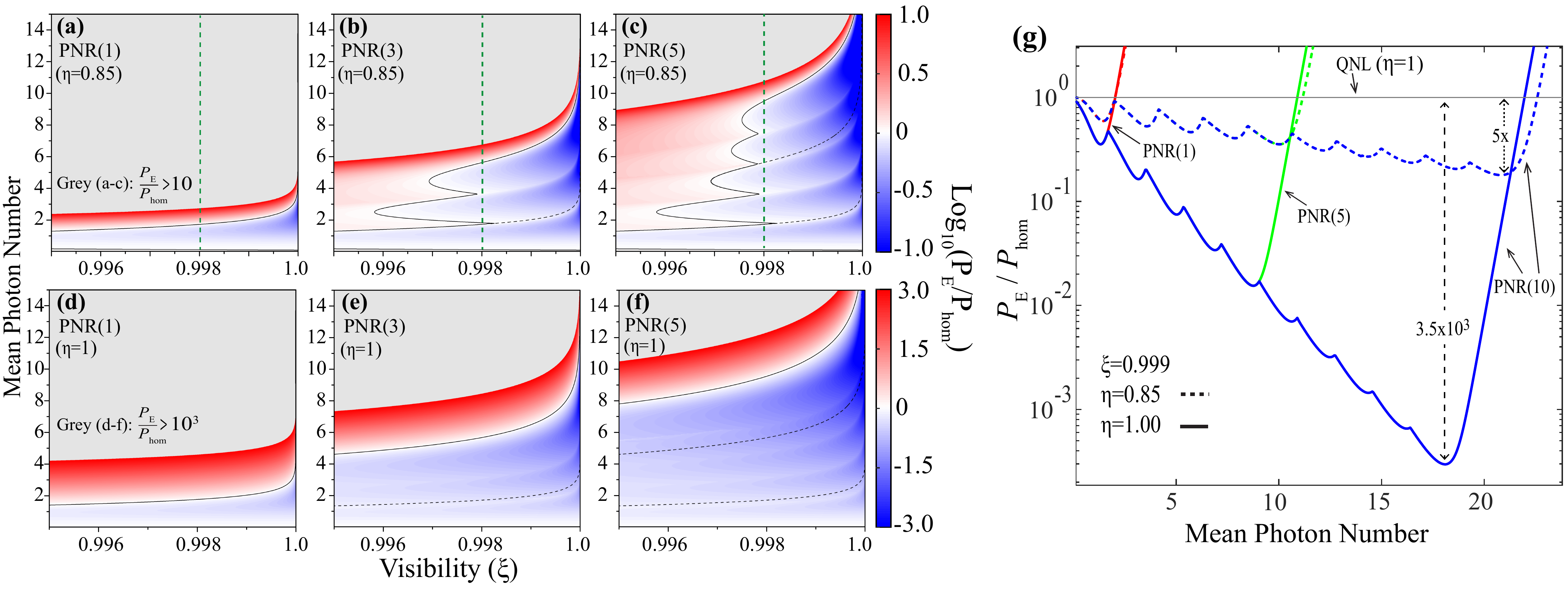}
\caption{\label{A} \textbf{Improvement of the PNR strategy over the QNL.} (a-f) Ratio of $P_\mathrm{E}(m)$ over the QNL in a logarithmic scale in the presence of noise resulting in reduced visibility $\xi$ as a function of $\xi$ and $|\alpha|^2$, for PNR($m$) for $m=1,3,5$ for cases $\eta = 0.85$ (a-c) and $\eta = 1$ (d-f). Note the different axis ranges in the top and bottom rows for $\eta = 0.85$  and $\eta = 1$ cases, respectively. The $\eta = 0.85$ case in (a-c) corresponds to the expected performance of our system with losses of about 12$\%$ using a superconducting PNR detector with $\eta=0.98$ \cite{fukuda11, lamas13}. Blue regions correspond to points in $\xi$ and $|\alpha|^2$ surpassing the QNL where $P_\mathrm{E}<P_\mathrm{hom}$, while red regions correspond to points where $P_\mathrm{E}>P_\mathrm{hom}$. Grey regions correspond to ratios $P_\mathrm{E}/P_\mathrm{hom}>10$ and $P_\mathrm{E}/P_\mathrm{hom}>1000$ for cases $\eta = 0.85$  and $\eta = 1$, respectively. Green dashed lines in (a-c) with $\eta = 0.85$ indicate the expected performance with our experimental visibility of $\xi$=0.998. (f) Attainable improvement over the QNL with resolutions PNR(1), PNR(5), and PNR(10) with a realistic visibility of $\xi=0.999$ for cases $\eta=0.85$ (dashed lines) and $\eta=1$ (solid lines). PNR(10) achieves a reduction of error compared to the QNL of about 5 times at $|\alpha|^2=21$ for $\eta=0.85$, and about $3.5\times10^{3}$ times at $|\alpha|^2=18$ for $\eta=1$. Larger improvements are expected for higher $m$ and higher visibilities.}
\end{figure*}

Phase stability of the interferometer is achieved with a feedback loop in a 33\% duty cycle with a frequency stabilized 780 nm laser, a differential detector (DD) and a piezo (PZT) on the back of a mirror \cite{ferdinand17}. Our experiment achieves a system detection efficiency $\eta = 0.72(1)$, visibility $\xi = 0.998(1)$, dark count rate $\nu = 3.6\times10^{-3}(2)$ and an afterpulsing probability $P_\mathrm{AP} =1.10\times10^{-2}(2)$.

\emph{Results and Discussion.\textemdash}
Figure \ref{ExpResults} shows the experimental results for the single-shot discrimination strategy with photon number resolving detection PNR($m$) for $m=1,2,3,4$, in solid circles, with error bars representing 1-$\sigma$ statistical standard deviation. The colored dashed lines show the theoretical predictions with detection efficiency $\eta = 0.72$, visibility $\xi = 0.998$, dark count rate $\nu = 3.6\times10^{-3}$, and afterpulsing probability $P_\mathrm{AP} = 1.10\times10^{-2}$. Included are the ideal homodyne limit at the QNL, the QNL for the same detection efficiency as our implementation, $\eta = 0.72$, and the Helstrom bound. We observe that while the strategy with PNR(1) surpasses the adjusted QNL with $\eta = 0.72$, it only does so up to a mean photon number of $|\alpha|^2\approx2$. On the other hand, strategies with photon number resolution PNR($m$) become robust to realistic noise and imperfections, enabling discrimination below the adjusted QNL for larger $|\alpha|$ by increasing the photon number resolution PNR($m$). The comparison with homodyne detection in a system with $\eta = 0.72$ allows us to investigate the performance of the robust PNR strategy under noise and imperfections including non-ideal visibility and dark counts and excluding the effect of detection efficiency. However, we note that state-of-the-art homodyne detectors have near-unity detection efficiency \cite{andersen16, vahlbruch16}. In the Supplemental Material \cite{Supplemental} we include comparisons with state-of-the-art detectors with and without considering system losses.

We investigated the expected advantages of PNR strategies over the QNL under realistic conditions with different visibilities. Figure \ref{A}(a-f) shows the ratio of the error probability to the QNL (homodyne limit) in a logarithmic scale for strategies with PNR($1$), PNR($3$), and PNR($5$) as a function of visibility $\xi$ and mean photon number $|\alpha|^2$ for detection efficiencies $\eta=0.85$ (a-c) and $\eta=1$ (d-f). The case $\eta=0.85$ corresponds to the expected performance of our system with losses of about 12\% and using a superconducting PNR detector with detection efficiency of $\eta=0.98$ \cite{fukuda11, lamas13} and negligible dark counts. We observe that even in a realistic case with $\eta=0.85$ and moderate visibility it is possible to achieve discrimination below the ideal QNL, blue regions with $P_\mathrm{E}(m)<P_\mathrm{hom}$. Moreover, increasing PNR($m$) significantly extends the regions in the parameter space of $|\alpha|^2$ and $\xi$ at which discrimination below the QNL can be achieved. Solid black lines show the boundary $P_\mathrm{E}(m)=P_\mathrm{hom}$. Dashed black lines mark the boundaries for strategies with smaller photon number resolution, which shows that discrimination strategies with higher PNR($m$) achieve higher improvements over the QNL. Figure \ref{A}(g) shows the attainable reduction of error rate compared to the QNL for PNR(1), PNR(5), and PNR(10) \cite{fukuda11, gerrits12} expected with a realistic visibility of $\xi$=0.999 for cases $\eta=0.85$ and $\eta=1$.
We observe a substantial improvement over the QNL with PNR(10) of about $5$ times at $|\alpha|^2=21$ for $\eta$=0.85, and $3.5\times10^3$ times at $|\alpha|^2=18$ for $\eta$=1. Moreover, this improvement increases with number resolution beyond 10 photons at higher optical energies, which is achievable with current PNR technologies \cite{gerrits12}.

Additional numerical studies included in the Supplemental Material \cite{Supplemental} show that PNR does not only provide robustness to noise resulting in reduced visibility, but also to other sources of imperfections, such as dark counts of non-ideal detectors. Moreover, our studies indicate that PNR detection, when used in intensity-modulated alphabets such as on-off-keying, also provides robustness to dark counts. These optimized PNR discrimination strategies complement the work in measurements with PNR for two coherent states used to assist quantum key distribution \cite{wittmann10}, and to allow for phase monitoring \cite{bina16} and for increasing the information extracted from a measurement \cite{han17}.

\emph{Conclusion.\textemdash}
We investigate and experimentally demonstrate a robust strategy for the discrimination of two nonorthogonal coherent states with minimum error based on single-shot optimized measurements with photon number resolving (PNR) detection. This PNR strategy generalizes near-optimal single-shot strategies of coherent states by increasing the number resolution to enable discrimination below the QNL under realistic conditions with noise and imperfections. Our experimental demonstration shows that this PNR strategy provides robustness to system nonidealities, and allows for surpassing the QNL adjusted for our system detection efficiency. Moreover, this PNR strategy also provides robustness to non-idealities of detectors such as dark counts for both phase encoding and intensity encoding schemes. Due to its robustness and the simplicity of single-shot measurements, this new strategy is inherently compatible with high-bandwidth communication technologies while performing below the QNL under realistic conditions. We expect that our work will motivate further developments in fast PNR detectors and in applications of these robust measurements in quantum information and communications.

\begin{acknowledgments}
We thank Konrad Banaszek for fruitful discussions. This work was supported by the National Science Foundation (NSF) (PHY-1653670, PHY-1521016).
\end{acknowledgments}


\end{document}




\title{Robust Measurement for the Discrimination of Binary Coherent States - Supplemental Material}

\author{M. T. DiMario}

\affiliation{Center for Quantum Information and Control, Department of Physics and Astronomy, University of New Mexico,
Albuquerque, New Mexico 87131}

\author{F. E. Becerra}

\affiliation{Center for Quantum Information and Control, Department of Physics and Astronomy, University of New Mexico,
Albuquerque, New Mexico 87131}
\email{fbecerra@umd.edu}

%
%

\maketitle

\section{Error probability}

The PNR discrimination strategy uses the maximum \emph{a posteriori} probability (MAP) criterion for the decision procedure. The probability of success for the discrimination of two coherent states $|\alpha\rangle$ and $|-\alpha\rangle$ is:
\begin{equation}
P_{C}=P(\alpha)P_{c}(\alpha)+P(-\alpha)P_{c}(-\alpha)
\label{PC}
\end{equation}
where $P(\pm\alpha)$ is the prior probability for the input state and $P_{c}(\pm\alpha)$ is the probability of correct discrimination given the input state was $|\pm\alpha\rangle$.

For a displaced photon number detection measurement, the MAP strategy assumes that given that we have observed $k$ photons and the input state was displaced by $\beta$, the correct state is $|\alpha\rangle$ if:
\begin{equation}
P(\alpha|k,\beta)>P(-\alpha|k,\beta),
\label{MAPalpha}
\end{equation}
but the correct state is $|-\alpha\rangle$ if:
\begin{equation}
P(\alpha|k,\beta)<P(-\alpha|k,\beta).
\label{MAP_minus_alpha}
\end{equation}

It follows that for the MAP decision strategy, the probability of correct discrimination $P_{c}(\alpha)$ for state $|\alpha\rangle$ is:
\begin{equation}
P_{c}(\alpha)=\langle\alpha|\Big(\sum_{k}\hat{\Pi}_{k}\Big[P(\alpha|k,\beta)>P(-\alpha|k,\beta)\Big]_{C}\Big)|\alpha\rangle
\label{Pc_alpha}
\end{equation}
where the symbol $[...]_{C}$ indicates the application of the MAP criterion and equals 1 if the condition inside the brackets is satisfied, and zero otherwise. Then, only the terms in the sum over $k$ for which the condition in Eq. (\ref{MAPalpha}) is satisfied contribute to $P_{c}(\alpha)$. $\hat{\Pi}_{k}=\hat{D}^\dag(\beta)|k\rangle\langle k|\hat{D}(\beta)$ is the displaced photon-number-resolving detection operator.

Given that $\langle\alpha|\hat{\Pi}_{k}|\alpha\rangle=|\langle k|\hat{D}(\beta)|\alpha\rangle|^2=P(k|\alpha,\beta)$ is the conditional probability for detecting $k$ photons given the input state $|\alpha\rangle$ and displacement field $\beta$, Eq. (\ref{Pc_alpha}) becomes
\begin{equation}
P_{c}(\alpha)=\sum_{k}P(k|\alpha,\beta)\Big[P(\alpha|k,\beta)>P(-\alpha|k,\beta)\Big]_{C}
\end{equation}

Similarly, the probability of correct discrimination $P_{c}(-\alpha)$ for state $|-\alpha\rangle$ is:
\begin{equation}
P_{c}(-\alpha)=\sum_{k}P(k|-\alpha,\beta)\Big[P(\alpha|k,\beta)<P(-\alpha|k,\beta)\Big]_{C}
\end{equation}

Using Bayes' theorem for equiprobable input states, $P(\alpha)=P(-\alpha)=\frac{1}{2}$, the condition in Eq. (\ref{MAPalpha}) is equivalent to:
\begin{equation}
P(k|\alpha,\beta)>P(k|-\alpha,\beta)
\end{equation}
Similarly, the condition in Eq. (\ref{MAP_minus_alpha}) is equivalent to
\begin{equation}
P(k|\alpha,\beta)<P(k|-\alpha,\beta).
\end{equation}

It follows that the total probability of correct discrimination in Eq. (\ref{PC}) for this strategy is:
\begin{eqnarray}\nonumber
P_{\textrm{C}}&=&\frac{1}{2}\sum_{k}\Big\{P(k|\alpha,\beta)\Big[P(k|\alpha,\beta)>P(k|-\alpha,\beta)\Big]_{C}+\\ \nonumber
&&P(k|-\alpha,\beta)\Big[P(k|\alpha,\beta)<P(k|-\alpha,\beta)\Big]_{C}\Big\}\\
&&=\frac{1}{2}\sum_{k}\max_{\pm\alpha}\{P(k|\pm\alpha,\beta)\}
\end{eqnarray}

where the $\max\{\}$ is taken over the input states ${|\pm\alpha\rangle}$. Therefore the probability of error for a PNR strategy that can discriminate up to $m$ photons, $k=0,1,2,...,m$, is:
\begin{equation}
P_{\textrm{E}}=1-P_{\textrm{C}}=1-\frac{1}{2}\sum_{k=0}^{m}\max_{\pm\alpha}{P(k|\alpha,\beta)}.
\end{equation}


\section{Optimal displacement Field}

\begin{figure}
\centering\includegraphics[width=8.5 cm]{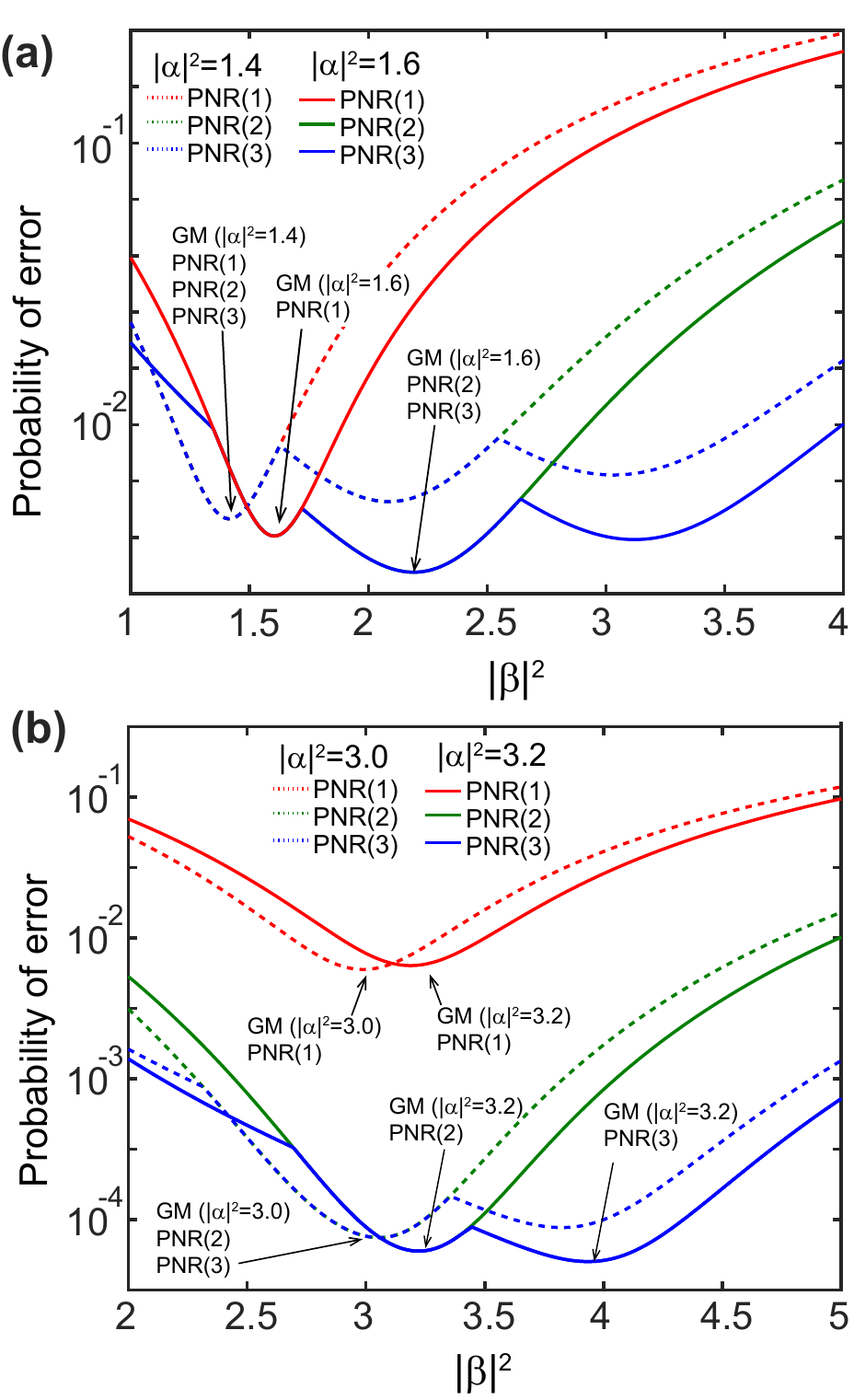}
\caption{\label{A} \textbf{Probability of error as a function of displacement $\beta$.} The probability of error as a function of displacement $\beta$ for mean photon numbers $|\alpha|^2$ (a) 1.4 and 1.6, and (b) 3.0 and 3.2, for strategies with different photon number resolutions PNR($m$) with $m=1,2,3$. Note the change in global minima for different PNR($m$) for different $m$ as $|\alpha|^2$ changes from 1.4 to 1.6 in (a) and from 3.0 to 3.2 in (b). GM; global minimum for different PNR($m$) strategies.
}\label{Pe-beta}
\end{figure}

The optimal displacements $\beta_{opt}$ are calculated by minimizing the probability of error $P_{E}(\alpha,\beta, m)$ in Eq. (3) in the main manuscript, for a given PNR($m$) strategy with number resolution $m$ and a given mean photon number $|\alpha|^2$. $P_{E}(\alpha,\beta, m)$ as a function of displacement field $\beta$, for a given PNR($m$) and $|\alpha|$, is a function with multiple minima (see Fig. S2). For a strategy with number resolution $m$, the number of minima is about $m$. The minimization of $P_{E}(\alpha,\beta, m)$, requires finding the global minimum (GM) among these minima. Fig. S2 shows $P_{E}(\alpha,\beta, m)$ as a function of $\beta$ for four mean photon numbers: Fig. S2(a) corresponds to $|\alpha|^2$ of 1.4 and 1.6; and Fig. S2(b) corresponds to $|\alpha|^2$ of 3 and 3.2. These values are around the values of $|\alpha|^2$ at which discrete jumps in $\beta_{opt}$ are observed  in Figure 1(b) in the main manuscript for different number resolutions $m$.

We observe in Fig. S2(a) that at $|\alpha|^2=1.4$, PNR(1), PNR(2), and PNR(3) have the same global minimum at about $|\beta|^2=1.4$. However, for $|\alpha|^2=1.6$, while the global minimum for PNR(1) is at about $|\beta|^2=1.6$, for PNR(2) and PNR(3) the global minimum now is at $|\beta|^2=2.3$. In a similar way, we observe in Figure S2(b) that while for $|\alpha|^2=3$, the global minima for PNR(2) and PNR(3) are the same at about $|\beta|^2=3.1$, when the mean photon number increases to $|\alpha|^2=3.2$, we find that the global minimum for PNR(2) locates at $|\beta|^2=3.2$, but the new global minimum for PNR(3) is at $|\beta|^2=3.9$. These changes of global minima in $P_{E}(\alpha,\beta, m)$ for strategies with different PNR($m$) make the optimal displacements $\beta_{opt}$ show jumps around these values of $|\alpha|^2$ that can be observed in Figure 1(b) in the main manuscript for different photon number resolutions $m$.


\section{Probability of After-pulsing and Experimental Runs}

\textbf{After-pulsing.} After-pulsing modifies the probability of registering different numbers of counts and impacts the PNR strategy. With after-pulsing, the probability of two photon detections $P_{2}$ becomes $P_{2} + P_\mathrm{AP}P_{1}$. The probability of three photon detections $P_{3}$ becomes $P_{3} + 2P_\mathrm{AP}P_{2} + P_\mathrm{AP}^{2}P_{1}$, and similarly for higher photon detections. We determined the after-pulsing probability of the APD in our experiment to be $P_\mathrm{AP} =1.10\times10^{-2}$ by using time-delayed photon counting measurements. This information is integrated in our simulations of the expected experimental results shown in Fig. (3) in the main manuscript.
\\
\\
\textbf{Experimental runs.} Each experimental run consisted of different numbers of independent experiments for different power ranges and PNR($m$) strategies. Higher powers result in lower errors and require more experiments. For PNR(1), one run consists of $1\times10^{5}$ experiments.
For PNR(2), one run requires from $1\times10^{5}$ to $5\times10^{5}$ experiments.
For PNR(3), one run requires from $1\times10^{5}$ to $3\times10^{6}$ experiments.
And for PNR(4), one run requires from $1\times10^{5}$ to $8\times10^{6}$ experiments.


\section{Performance comparisons}

\begin{figure}
\centering\includegraphics[width=8.5 cm]{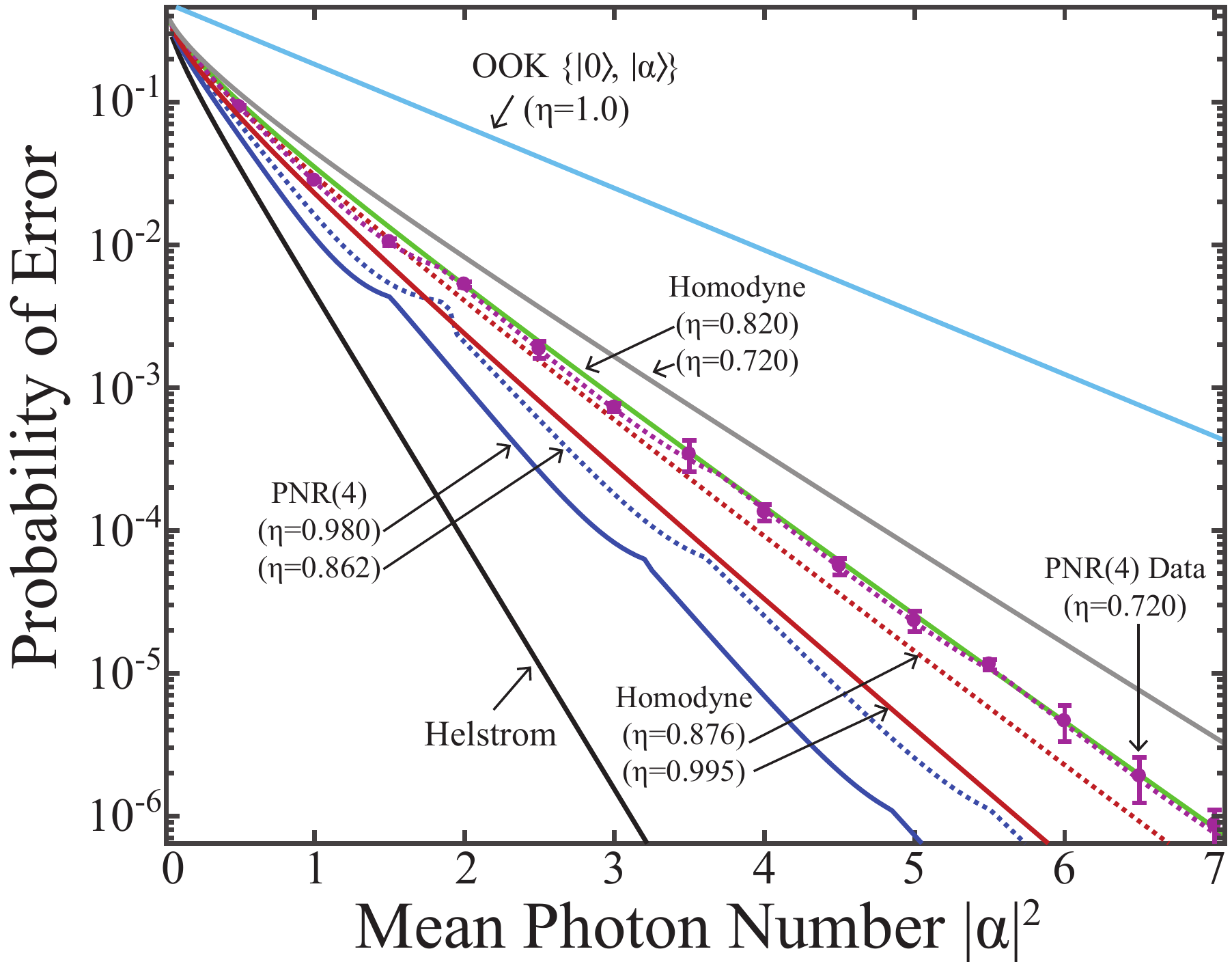}
\caption{\label{A} \textbf{Comparison of the PNR strategy with state-of-the-art detectors.}
PNR(4) strategy in our experiment with overall detection efficiency of $\eta=72\%$ resulting from the APD detector efficiency $\eta_{\textrm{APD}}=82\%$ and the system efficiency of $\eta_{\textrm{sys}}=88\%$, and with a in visibility $\xi=0.998$: data (magenta circles) and simulation (magenta dashed line). State-of-the-art homodyne detection \cite{andersen16, vahlbruch16} with $\eta_{\textrm{hom}}=99.5\%$ without system losses (red solid line) and when considering losses with system efficiency of $\eta_{\textrm{sys}}=88\%$ (red dashed line).  State-of-the-art PNR detector with detection efficiency $\eta_{\textrm{PNR}}=98\%$ \cite{fukuda11, lamas13} without system losses (blue solid line) and when considering losses with system efficiency $\eta_{\textrm{sys}}=88\%$ (blue solid line). Homodyne detector with a system efficiency of $\eta=82\%$ (green solid line) performs similar to a PNR(4) with our system with overall detection efficiency of $\eta=72\%$  and visibility $\xi=0.998$. On-off keying (OOK) intensity modulated alphabet with $\{|0\rangle,|\alpha\rangle\}$ with perfect efficiency and no noise.
}\label{PerformanceComparison}

\end{figure}

We compare the performance of the PNR strategy under realistic noise and imperfections
with state-of-the-art homodyne detectors \cite{andersen16, vahlbruch16} with and without considering system losses. In our experiment, the noise and imperfections result in a visibility of $\xi=0.998$. The total efficiency in our experiment $\eta=72\%$ results from the detector efficiency of the APD, $\eta_{\textrm{APD}}=82\%$, and the system efficiency of $\eta_{\textrm{sys}}=88\%$ resulting from losses in the setup. Figure (S1) shows the performance of our experiment compared to state-of-the-art homodyne detection with efficiency $\eta_{\textrm{hom}}=99.5\%$ \cite{andersen16, vahlbruch16} considering a lossless system (red solid line) and a system with losses equal to our system $\eta_{\textrm{sys}}=88\%$ (red dashed line). Figure (S1) also includes the expected performance of the PNR strategy when using state-of-the-art PNR detectors \cite{andersen16, vahlbruch16} with $\eta_{\textrm{PNR}}=98\%$, with noise and imperfections that result in a visibility of $\xi=0.998$ in a lossless system (blue solid line) and a system with losses $\eta_{\textrm{sys}}=88\%$ (blue dashed line). We note that homodyne detection with 99.5\% detection efficiency with no losses is similar to the ideal QNL. We note that a homodyne detection used with a system with total loss of about 18\%, corresponding to an efficiency $\eta=82\%$ (green solid line), would perform similar to a PNR(4) in our experiment with total system efficiency of $\eta=72\%$ and $\xi=0.998$.

Figure S1 also shows the performance of an intensity-modulated alphabet with $\{|0\rangle,|\alpha\rangle\}$ such as on-off-keying (OOK) as a light blue line.This alphabet is not affected by errors associated with non-unity visibility but has a sensitivity that is lower than coherent-encoded alphabets. We note that while OOK schemes are not affected by reduced visibility, these schemes are very sensitive to detector dark counts. The section \emph{PNR ROBUSTNESS TO DARK COUNTS} below shows how PNR detection can provide robustness to detector dark counts for both phase encoded and intensity encoded OOK schemes.


\section{PNR robustness to dark counts}

\begin{figure}
\centering\includegraphics[width=8.5 cm]{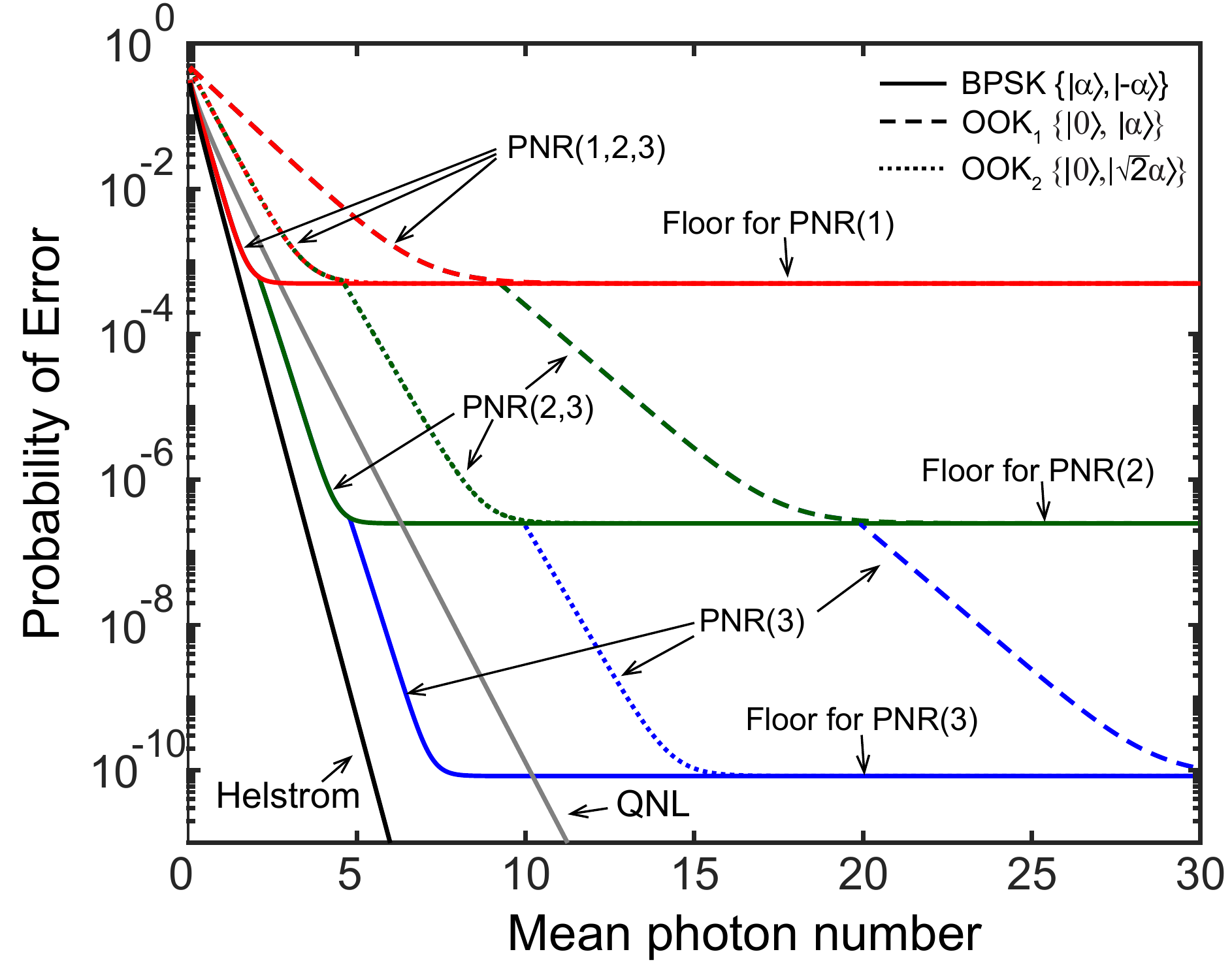}
\caption{\label{A} \textbf{PNR discrimination strategy with dark counts.} Probability of error as a function of mean photon number for PNR($m$) strategies with  $m=1,2,3$ with detector dark count probability of $10^{-3}$ for three alphabets: binary-phase-shift keying $\{|\alpha\rangle,|-\alpha\rangle\}$; on-off keying with the same peak power $\{|0\rangle,|\alpha\rangle\}$ (OOK$_1$); and on-off keying with the same average power $\{|0\rangle,|\sqrt{2}\alpha\rangle\}$ (OOK$_2$). PNR strategies with BPSK provide higher sensitivities than for OOK. Note that for all the input alphabets, BPSK and the two OOKs, the error floor due to dark counts is the same for a given photon number resolution PNR($m$).
}\label{Pe-beta}
\end{figure}

We have investigated the PNR strategy in situations with detectors having dark counts, which  cannot be captured by the reduction of interference visibility. We have observed that the PNR strategy provides robustness to detector dark counts for both phase encoding and intensity encoding schemes. Figure S3 shows the probability of error for the discrimination of two states with PNR($m$) strategies with $m=1,2,3$ assuming a dark count rate of $10^{-3}$ dark counts per pulse, for three possible alphabets: binary-phase shift keying  $\{|\alpha\rangle,|-\alpha\rangle\}$; on-off keying (OOK) with alphabet  $\{|0\rangle,|\alpha\rangle\}$ (OOK$_1$); and OOK with alphabet $\{|0\rangle,|\sqrt{2}\alpha\rangle\}$ (OOK$_2$). OOK$_1$ corresponds to an OOK scheme with the same peak power as the BPSK scheme, and OOK$_2$ corresponds to a scheme with the same average power as the BPSK scheme. Note that BPSK provides an overall higher sensitivity than the OOK schemes, reaching lower probability.

We observe that PNR detection provides robustness against dark counts to all these alphabets, allowing to mitigate somewhat the effects of dark counts. Moreover, higher PNR($m$) provides higher levels of robustness.
We note that the rate of dark counts sets a noise floor for the probability of error, which is the same for different alphabets, and that increasing the PNR resolution by one reduces this error floor by about 3 orders of magnitude.

\bibliography{Base}